\documentclass[prd, aps, superscriptaddress, preprintnumbers, twocolumn, floatfix, nofootinbib]{revtex4}

\usepackage{amsfonts}
\usepackage{amsmath}
\usepackage{amssymb}
\usepackage{bm}
\usepackage{dcolumn}
\usepackage{graphicx}   
\usepackage[latin1]{inputenc}
\usepackage{latexsym}
\usepackage{rotating}
\usepackage{hyperref}
\usepackage{graphicx}
\usepackage{color}

\newcommand\be{\begin{equation}}
\newcommand\ba{\begin{eqnarray}}
\newcommand\ee{\end{equation}}
\newcommand\ea{\end{eqnarray}}

\begin{document}

\title{\textbf{Axion Monodromy Inflation, Trapping Mechanisms and the Swampland}}

\author{Weijie Jin}
\email{wjin@student.ethz.ch}
\affiliation{Institute of Theoretical Physics, ETH Z\"urich, CH-8093 Z\"urich, Switzerland}

\author{Robert Brandenberger}
\email{rhb@physics.mcgill.ca}
\affiliation{Department of Physics, McGill University, Montr\'{e}al, QC, H3A 2T8, Canada and
Institute of Theoretical Physics, ETH Z\"urich, CH-8093 Z\"urich, Switzerland}

\author{Lavinia Heisenberg}
\email{laviniah@phys.ethz.ch}
\affiliation{Institute of Theoretical Physics, ETH Z\"urich, CH-8093 Z\"urich, Switzerland}

\date{September 2020}

\begin{abstract}

We study the effects of particle production on the evolution of the inflaton field in an axion monodromy model with the goal of discovering in which situations the resulting dynamics will be consistent with the {\it swampland constraints}. In the presence of a modulated potential the evolving background field (solution of the inflaton homogeneous in space) induces the production of long wavelength inflaton fluctuation modes. However, this either has a negligible effect on the inflaton dynamics (if the field spacing between local minima of the modulated potential is large), or else it traps the inflaton in a local minimum and leads to a graceful exit problem. On the other hand, the production of moduli fields at enhanced symmetry points can lead to a realization of {\it trapped inflation} consistent with the swampland constraints, as long as the coupling between the inflaton and the moduli fields is sufficiently large.

\end{abstract}

\maketitle

\section{Introduction}

The inflationary scenario \cite{inflation} has become the paradigm of early universe cosmology. However, it is unclear whether cosmological inflation can be embedded in an ultraviolet complete theory. Inflation is usually obtained using slow rolling on an almost constant potential in the context of an effective field theory of a canonically normalized scalar field $\phi$ minimally coupled to Einstein gravity. Recently, it has been argued that any such effective field theory which is consistent with superstring theory must satisfy a number of conditions (which are called the {\it swampland constraints}) \cite{Brennan, Palti}. The first condition is the {\it distance conjecture} \cite{Vafa1} which states that any effective field theory has an applicability range which is limited by $|\Delta \phi| < c_1 m_{pl}$, where $m_{pl}$ is the four space-time dimensional Planck mass, and $c_1$ is a coefficient of the order $1$. The second condition ({\it de Sitter conjecture}) states \cite{Vafa2, Krishnan} that the potential is either sufficiently steep or sufficiently tachyonic, i.e.
\ba \label{deSitter}
|\frac{V^\prime m_{pl}}{V}| \, &>& \, c_2 \,\,\,\,\, {\rm{or}} \nonumber \\
\frac{V^{\prime \prime} m_{pl}^2}{V} \, &<& \, -c_3 
\ea
where $c_2$ and $c_3$ are positive coefficients of the order $1$. These conditions effectively rule out standard single field slow-roll models of inflation \cite{noinflation} \footnote{These criteria also constrain dark energy models constructed using quintessence fields, but the current data do not yet rule them out \cite{Lavinia}. They also constrain significantly dark energy models with derivative self-interactions \cite{heisenberg2019horndeski}.}. They also rule out false vacuum models of inflation constructed using the effective field theory approach, while still leaving room for non-perturbative constructions of metastable de Sitter space such as those discussed recently in \cite{Dvali, Keshav}.

While the swampland criteria severely constrain standard slow-roll inflation, they can be consistent with other approaches to inflation such as warm inflation \cite{warm}, trapped inflation \cite{trapped} and chromo-natural inflation \cite{chromo}. In these scenarios, there are extra contributions to the equation of motion of the scalar field $\phi$ due to couplings with other matter fields which allow for accelerated expansion of space on a steep potential. Note, however, that these models continue to be in tension with the recently proposed {\it Trans-Planckian Censorship Conjecture} (TCC) \cite{Bedroya, Bedroya2}, as discussed in \cite{Vahid}. The latter constrains severely the future dynamics of many dark energy models \cite{heisenberg2020model}.

Before the swampland criteria gained attention, there had been many attempts to embed cosmological inflation into string theory (see e.g. \cite{Liam} for a textbook treatment). Axion monodromy inflation \cite{Eva1, Eva2} is considered to be one of the most promising scenarios. In this approach, a large effective field range is rendered consistent with the sub-Planckian range of the size of the extra spatial dimensions by a monodromy effect. An intuitive way to visualize this \cite{Eva1} is to consider $\phi$ to be related to the length of a string which connects a bulk brane to the brane we live on, and having the string wind the base manifold many times. The overall potential for $\phi$ is then modulated each time the brane positions come to overlap. Initially, axion monodromy inflation was suggested as a way to implement standard slow-roll large field inflation. Later on it was realized \cite{trapping} (see also \cite{Scott}) that at the brane crossing points the masses of other string degrees of freedom become small, and that these fields can then be produced by the rolling $\phi$ field by a mechanism similar to how matter can be produced in the preheating process at the end of inflation \cite{preheating}. The particle production creates a new effective friction term which can slow down the motion, as already shown in \cite{trapped, Vahid}. 

In this note we take a closer look at the effects of moduli field production at enhanced symmetry points. We show that particle production via the intrinsic nonlinearities in the scalar field equation of motion produced by the modulation of the potential is insufficient to render axion monodromy inflation consistent with the swampland conjectures - the effect of $\phi$ particle production is either negligible, or else it traps $\phi$ in a false vacuum state, leading to a graceful exit problem. Moduli field production, on the other hand, leads to sufficient effective friction and allows for accelerated expansion for sub-Planckian field values for which the swampland criteria are satisfied, without trapping $\phi$ in a false vacuum. 
 
In the following section we will consider $\phi$ particle production due to the nonlinear terms in the $\phi$ equation of motion generated by the modulation of the $\phi$ potential. We show that, depending on the amplitude of the modulation term, $\phi$ is either trapped in one of the false vacua it encounters, or else the effect of $\phi$ particle production is negligible.
 
From the point of view of string theory, however, it is not consistent to just consider the modulation of the potential for $\phi$ but to neglect the effect of the rolling $\phi$ on the moduli fields which become light at the enhanced symmetry points. In Section 3 we hence study, similar to what was done in \cite{trapped}, moduli production at these points, making use of the analysis of \cite{trapping}. We find that for sufficiently large values of the coupling between $\phi$ and the moduli fields, moduli production slows down the motion of $\phi$ sufficiently such that accelerated expansion can occur for field values $|\phi| \leq m_{pl}$ for which the swampland criteria are obeyed: the relative slope of the potential is sufficiently large, and super-Planckian field values are not required. Thus, both the distance and de Sitter criteria are obeyed.
 
We will be using natural units in which the speed of light and Planck's constant are set to $1$. We analyze the evolution of the matter fields in a background cosmological model with scale factor $a(t)$, in terms of which the Hubble expansion rate is $H(t) = {\dot{a}}/a$, where $t$ is the physical time and an overdot indicates a derivative with respect to $t$. The coordinates ${\bf{x}}$ are comoving spatial coordinates, and ${\bf{k}}$ are the associated comoving momenta, with $k$ denoting the magnitude. The reduced Planck mass is denoted by $m_{pl} \equiv [1/(8 \pi G)]^{1/2}$.

\section{Particle Production from the Modulation of the Potential}

\subsection{Model}

We consider a particular version of the axion monodromy inflation model in which the axion is described as a scalar field $\phi$ with a linear potential which is modulated by a sinusoidal correction term induced by instantons \cite{Eva2}
\begin{equation}\label{potential}
    V(\phi) \, = \, \mu^3 \phi + \Lambda^4 \cos(\frac{\phi}{2\pi f}) \, ,
\end{equation}
where $\mu$ and $\Lambda$ are model-dependent parameters determined by the brane  compactification and the instanton action respectively, satisfying $\Lambda\ll \mu$, while $f$ characterizes the frequency of the periodic correction, and $f \ll m_{pl}$.

Assuming that the axion is the inflaton, and neglecting the effect of the sinusoidal term, the inflaton must start at $\phi \sim 11 m_{pl}$ in order to obtain 60 e-foldings of accelerated expansion \cite{Eva2}. However, such field values are obviously in tension with the distance conjecture, and also the de Sitter conjecture, since for the linear potential of the axion the first line of the condition (\ref{deSitter}) demands that
\begin{equation}\label{desitterconstrain}
    \phi \, \lesssim \, 1 m_{pl} \, ,
\end{equation}
setting the constant $c_2 = 1$, and is thus in contradiction with the requirement from 60 e-foldings of inflation.

However, due to the nonlinearities in the equation of motion for $\phi$ induced by the periodic correction to the potential, there will be production of $\phi$ particles, and this particle production effect will back-react and slow down the evolution of $\phi$. In the following, we will consider $\phi$ particle production and its effect on the evolution of the background value of $\phi$, in order to examine whether this effect is sufficient to cause slow-roll inflation with field values which are consistent with the de Sitter constraint.

\subsection{First Order Perturbation of the Axion Evolution}

Using the axion potential eq. (\ref{potential}) in the scalar field action, the equation of motion of the spatially homogeneous background solution $\phi_0(t)$ is
\begin{equation}\label{background}
        \ddot{\phi_0} + 3H\dot{\phi_0} + \mu^3 
        - \frac{\Lambda^4}{2\pi f}\sin\frac{\phi_0}{2\pi f} \, = \, 0 \, 
\end{equation}
where the Hubble expansion rate is given by
\be        
       H \, = \, \frac{1}{\sqrt{3}m_{pl}}\sqrt{\frac{1}{2}\dot{\phi_0}^2+V(\phi_0)} \, .
\ee

We will study the space-dependent $\phi$ fluctuations in first order perturbation theory which are induced by the nonlinearities. The ansatz for $\phi$ is
\begin{equation}\label{fullphi}
    \phi(\mathbf{x},t) \, = \, \phi_0(t)+\varphi(\mathbf{x},t) \, .
\end{equation}
Expanding the equation of motion for $\phi$ to first order in the amplitude of the fluctuations, and neglecting for a moment the expansion of space (we will include this effect later on), we obtain the following equation of motion for $\varphi$
\begin{equation}
    \ddot{\varphi} - \nabla^2 \varphi + V''(\phi_0)\varphi \, = \, 0.
\end{equation}
Since the equation of motion is linear, each Fourier mode $\varphi_k (t)$ evolves independently according to the equation
\begin{equation}
    \ddot{\varphi}_k + (k^2 - \frac{\Lambda^4}{(2\pi f)^2}\cos \frac{\phi_0}{2\pi f})\varphi_k
    \, = \, 0.
\end{equation}
This equation describes a harmonic oscillator with a time-dependent mass. The time-dependence of the mass can lead to particle production. Particle production is expected to be important if the adiabaticity condition on the effective frequency $\omega_k$ is violated, i.e.
\be \label{alpha}
\alpha \, \equiv \, \frac{d\omega^2/ dt }{\omega^3} \, > \, 1 \, .
\ee
For modes with 
\be
k \, \gg \, \frac{\Lambda^2}{2\pi f} \, ,
\ee
the adiabaticity condition is not violated, and we obtain oscillating behavior, no amplification and hence no particle production. Therefore, they do not effect the evolution of the background . 

In contrast, modes with
\be \label{range2}
k \, \ll \, \frac{\Lambda^2}{2\pi f}
\ee
can be amplified. This draws energy from the background field and can slow down the rolling. The typical time scale of the instability is expected to be shorter than the Hubble time step, thus justifying the neglect of the Hubble friction term. In the numerical parts of the following analysis, we use the parameter values $\mu\sim 0.1$, $\Lambda\sim 0.01$, $f\sim 0.001$, all in Planck units.

\begin{figure}[htbp]
\centering
\includegraphics[scale=0.4]{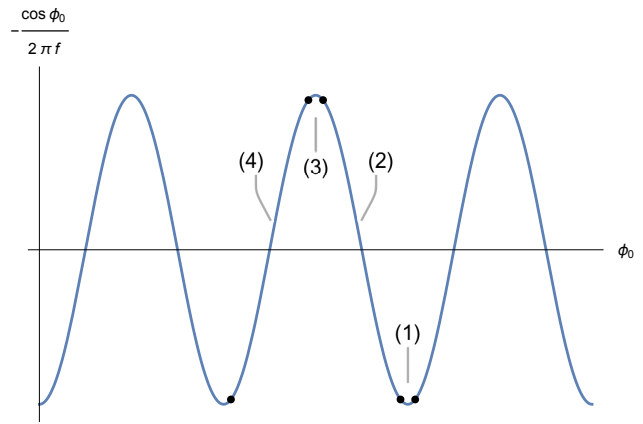}
\caption{The periodical term in the equation of motion of $\varphi_k$. $\phi_0$ decreases as time proceeds, and the evolution of $\varphi_k$ in each period is analysed separately for the phases (1),(2),(3),(4).}
\label{phipotential}
\end{figure}

For modes in the range (\ref{range2}), the approximate form of the mode equation becomes 
\begin{equation}\label{varphikevolution}
    \ddot{\varphi}_k-\frac{\Lambda^4}{(2\pi f)^2}\cos( \frac{\phi_0}{2\pi f})\varphi_k=0.
\end{equation}
The evolution of $\varphi_k$ in one period of $\phi_0$ can be approximated separately in four phases, as shown in Figure \ref{phipotential}. In each phase, the sinusoidal term is taken to be linear in $\phi_0$, and is hence also linear in time if we use the background solution $\phi_0(t)$.
The division of the phases is determined by the adiabaticity parameter $\alpha$ defined in (\ref{alpha}). We start with 
\begin{equation}
    \omega^2 \, \equiv \, -\frac{\Lambda^4}{(2\pi f)^2}\cos \frac{\phi_0}{2\pi f},
\end{equation}
and then $\alpha$ becomes
\ba
    \alpha \, &=& \, \frac{1}{\Lambda^2}\left|\frac{\sin(\frac{\phi_0}{2\pi f})\dot{\phi}_0}{\cos^{3/2}(\frac{\phi_0}{2\pi f})}\right| \\
    &\simeq& \,  \frac{1}{\Lambda^2}\left|\frac{\sin(\frac{\phi_0}{2\pi f})}{\cos^{3/2}(\frac{\phi_0}{2\pi f})}\right|\frac{m_{pl}}{\sqrt{3}}\mu^{3/2}\phi_0^{-1/2} \nonumber
\ea
where we made use of 
\be
\dot{\phi}_0 \, \simeq \, -\frac{\mu^3}{3H}
\ee
and 
\be
H \, \simeq \, \frac{1}{\sqrt{3}m_{pl}}\sqrt{\mu^3\phi_0} \, . 
\ee
The junctions between phases occur when $\alpha=1$, dividing the period according to whether $\omega^2$ changes rapidly or not.

In phases (1) and (3), the adiabaticity condition is satisfied ($\alpha\lesssim 1$), so that $\omega^2$ can be approximated with a constant $\omega^2\simeq \pm \frac{\Lambda^4}{(2\pi f)^2}$. However, the time duration of these phases is found to be much shorter than that of the whole period, and thus the $\varphi_k$ evolution in these two phases has negligible effects.

In phase (2), $\omega^2(t)$ can be expanded linearly around $\phi_0(t_2)$ where $\cos \frac{\phi_0(t_2)}{2\pi f}=0$
\begin{equation}\label{phase2}
    \omega^2 \, \simeq \, -\frac{\Lambda^4}{(2\pi f)^3}\dot{\phi}_0(t_2) (t-t_2).
\end{equation}
Substituting eq. (\ref{phase2}) into eq. (\ref{varphikevolution}) we get 
\begin{equation}
    \ddot{\varphi}_k+P (t-t_2)\varphi_k \, = \, 0,
\end{equation}
in which 
\be
P \, \equiv \, -\frac{\Lambda^4}{(2\pi f)^3}\dot{\phi_0}(t_2) \, \simeq \, \frac{\Lambda^4}{(2\pi f)^3}\frac{m_{pl}}{\sqrt{3}}\mu^{3/2}\phi_0^{-1/2}(t_2) \, .
\ee 
This equation has a solution
\ba
    \varphi_k \, &=& \, C_1 Ai(P^{1/3}e^{i\frac{\pi}{3}}(t-t_2))\nonumber \\
       &+& \, C_2 Bi(P^{1/3}e^{i\frac{\pi}{3}}(t-t_2)), 
\ea
where $Ai$ and $Bi$ are Airy functions, and $C_1$ and $C_2$ are constants. The solution can be further expanded around $t=t_2$, because $|P^{1/3}(t-t_2)|\lesssim 0.05$ given the values of the parameters which we use, and the Airy functions are approximately linear in this range:
\ba
    Ai(x) \, &=& \, \frac{1}{3^{2/3}\Gamma(\frac{2}{3})}-\frac{x}{3^{1/3}\Gamma(\frac{1}{3})}+\mathcal{O}(x^3), \nonumber \\
    Bi(x) \, &=& \, \frac{1}{3^{1/6}\Gamma(\frac{2}{3})}+\frac{3^{1/6}x}{\Gamma(\frac{1}{3})}+\mathcal{O}(x^3).
\ea
Therefore, $\varphi_k$ evolves linearly in phase (2). Similarly, in phase (4), it also has a linear behavior.

In summary, the evolution of $\varphi_k$ in one period is nearly linear in time, when the boundary conditions between phases are matched. Connecting the solutions in different periods, $\varphi_k$ is found to grow linearly during the whole process
\begin{equation}\label{linearvarphi}
    \varphi_k \, = \, \frac{1}{\sqrt{2k}}+\sqrt{\frac{k}{2}}t,
\end{equation}
with the initial conditions determined by assuming that the modes start out in their vacuum state, and using quantum vacuum fluctuation initial conditions
\ba
\varphi_k(0) \, &=& \, \frac{1}{\sqrt{2k}} \nonumber \\
\dot{\varphi}_k(0) \, &=& \, \sqrt{\frac{k}{2}} \, .
\ea

However, when we consider the $\varphi_k$ evolution during the whole time evolution instead of during a single period, the effects of the expansion of space cannot be neglected. The equation of motion becomes
\begin{equation}
    \ddot{\varphi}_k + 3H\dot{\varphi}_k - \frac{\Lambda^4}{(2\pi f)^2}\cos( \frac{\phi_0}{2\pi f})\varphi_k \, = \, 0 
\end{equation}
instead of eq. (\ref{varphikevolution}). The Hubble damping term can be removed by the following field redefinition \cite{KLS97}
\begin{equation}
    \Phi_k \, \equiv \, a^{\frac{3}{2}}\varphi_k \, .
\end{equation}
In terms of the new field variable, the equation becomes
\begin{equation}
    \ddot{\Phi}_k - \left(\frac{\Lambda^4}{(2\pi f)^2}\cos( \frac{\phi_0}{2\pi f})+\delta\right)\Phi_k \, = \, 0 \, ,
\end{equation}
in which $\delta:=\frac{9}{4}H^2+\frac{3}{2}\dot{H}$ is negligible compared to the other terms, so that the behavior of $\Phi_k$ is similar to the solution eq. (\ref{linearvarphi}), but with different initial conditions. Setting the initial time to be $t = 0$, and normalizing the scale factor as $a(t=0) = 1$, we get 
\ba
    \Phi_k(0) \, &=& \, \frac{1}{\sqrt{2k}} \\
     \dot{\Phi}_k(0) \,  &=& \, \frac{3}{2}H(0)\frac{1}{\sqrt{2k}}+\sqrt{\frac{k}{2}} \nonumber
\ea
so that
\begin{equation}
    \Phi_k \, = \, \frac{1}{\sqrt{2k}}+\left(\frac{3}{2}H(0)\frac{1}{\sqrt{2k}}+\sqrt{\frac{k}{2}}\right)t,
\end{equation}
and hence
\begin{equation}
    \varphi_k \, = \, \frac{a^{-\frac{3}{2}}}{\sqrt{2k}}+a^{-\frac{3}{2}}\left(\frac{3}{2}H(0)\frac{1}{\sqrt{2k}}+\sqrt{\frac{k}{2}}\right)t.
\end{equation}
Note that in a background with accelerated expansion, the dilution due to the expansion of space overwhelms the increase in the number of particles.
\\

\subsection{Back-reaction of the Perturbations}

At quadratic order in the amplitude of the fluctuations, these back-react on the homogeneous mode of $\phi$.  Expanding the equation of motion for $\phi$ in eq. (\ref{fullphi}) to second order and averaging over the fluctuation field $\varphi$ yields the equation
\ba \label{backreaction}
    \ddot{\phi_0} &+& 3H\dot{\phi_0} + \mu^3 \\
    &-& \frac{\Lambda^4}{2\pi f}\sin\frac{\phi_0}{2\pi f}+\frac{1}{2}\frac{\Lambda^4}{(2\pi f)^3}\sin \left(\frac{\phi_0}{2\pi f}\right)\langle\varphi^2\rangle \, = \, 0 \, . \nonumber
\ea
This equation differs from eq. (\ref{background}) in that it includes the effects of the back-reaction of the fluctuation field $\varphi$ on the spatial homogeneous solution $\phi_0$.

The expectation value $\langle\varphi^2\rangle$ can be found by integrating the contributions of all of the Fourier modes which are being excited, i.e. from $k = 0$ until the critical value $k=\frac{\Lambda^2}{2\pi f}$:
\begin{widetext}
\begin{align}\label{varphisquare}
    \langle\varphi^2\rangle&\simeq \int_0^{\frac{\Lambda^2}{2\pi f}}dk 4\pi k^2 \langle\varphi^2_k\rangle\nonumber\\
    &=\frac{4\pi}{a^3}\left[\frac{1}{8}\left(\frac{\Lambda^2}{2\pi f}\right)^4+\frac{9}{16}H^2(0)\left(\frac{\Lambda^2}{2\pi f}\right)^2+\frac{1}{2}H(0)\left(\frac{\Lambda^2}{2\pi f}\right)^3\right]t^2\nonumber\\
    &+\frac{4\pi}{a^3}\left[\frac{1}{3}\left(\frac{\Lambda^2}{2\pi f}\right)^3+\frac{3}{4}\left(\frac{\Lambda^2}{2\pi f}\right)^2H(0)\right]t
\end{align}
\end{widetext}
where we used eq. (\ref{backreaction}) and assumed $\langle\varphi^2\rangle|_{t=0}=0$. Substituting eq. (\ref{varphisquare}) into eq. (\ref{backreaction}), the $\phi_0$ evolution can be determined and compared with the background solution.

\subsection{Numerical Solutions of the $\phi_0$ Evolution}

In the following we numerically solve the equation (\ref{backreaction}) for the homogeneous field mode corrected by the back-reaction effects and compare the result with the solution $\phi_0$ of the background without any back-reaction terms. The background solution $\phi_0$ comes from solving eq. (\ref{background}). We have checked that the precise values of $\Lambda$ and $f$ have a negligible effect on our overall conclusions, as long as $\Lambda \ll \mu$.

The $\phi_0$ evolution including the back-reaction of the produced particles comes from solving eq. (\ref{backreaction}), using 
\begin{equation}
    H \, = \, \frac{1}{\sqrt{3}m_{pl}}\sqrt{\frac{1}{2}\dot{\phi_0}^2+V(\phi_0)},
\end{equation}
and
\begin{equation}
    a(t) \, = \, \exp \int_0^t H(t')dt'.
\end{equation}

\begin{widetext}
\begin{figure}[htbp]
\centering
\includegraphics[scale=1]{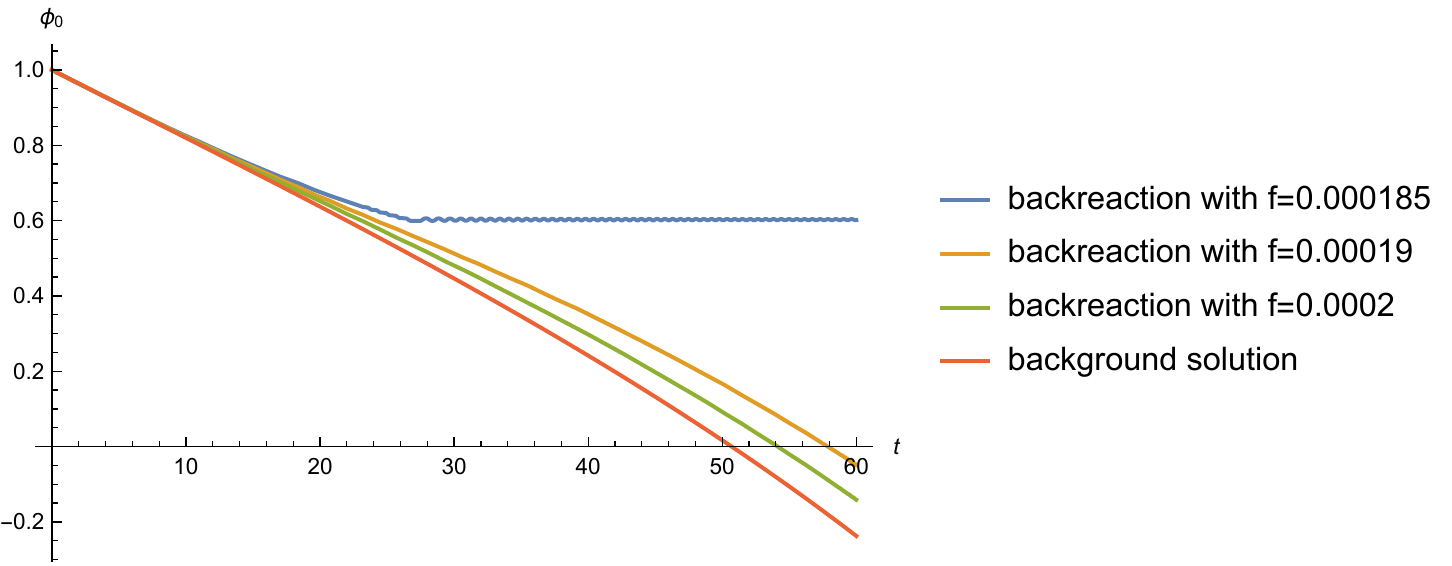}
\caption{The evolution of $\phi_0$ for different values of $f$, with fixed $\mu=0.1$, $\Lambda=0.01$, compared with the background solution. Both the field values and the time are in Planck units.}
\label{axionevolution}
\end{figure}
\end{widetext}

To check whether $\phi$ particle production can render the model consistent with the swampland conjectures, we start the evolution with the largest field value consistent with the distance conjecture, namely $\phi_0 = 1$ (in Planck units). Fixing $\mu$ and $\Lambda$, we find that the parameter $f$ influences the fate of $\phi_0$, as shown in Figure \ref{axionevolution}. If the period $f$ of the oscillations of the potential is small, then there is a big back-reaction effect and $\phi_0$ gets trapped almost immediately, leading to a graceful exit problem. On the other hand, if the local minima of the potential are well separated, then the effect of particle production is negligible, and no inflation can occur. As shown in the graph, the transition in the behavior of the solution as a function of $f$ is very sharp. 

\section{Including the Effects of Moduli Field Production}

\subsection{Model}

From the point of view of string theory, the model considered in the previous section is incomplete. It includes the modulation of the potential due to the monodromy effects, but it does not take into account that the monodromy points (the local minima of the modulated potential) are enhanced symmetry states at which a tower of string modes (which we call ``moduli'' in the following) become massless. If these fields couple to the inflaton, as is expected, then the rolling inflaton field will lead to the production of moduli quanta \cite{trapping}.

To study the effect of moduli production on the inflaton dynamics we consider the following Lagrangian \cite{trapped} 
\ba \label{trappedlagrangian}
    \mathcal{L} \, &=& \, \frac{1}{2}\partial_{\mu}\phi \partial^{\mu}\phi -V(\phi)  \\
    &+& \frac{1}{2}\sum_i (\partial_{\mu}\chi_i\partial^{\mu}\chi_i-g^2(\phi-\phi_i)^2\chi_i^2)
    \, , \nonumber
\ea
where the potential is assumed to be
\begin{equation}
    V(\phi) \, = \, \mu^3 \phi \, ,
\end{equation}
where we are neglecting the modulation of the potential. As discussed in the previous section, this is a good approximation as long as the minima are not too closely packed. This model describes the axion monodromy field coupled to the string degrees of freedom $\chi_i$, which become light periodically when the rolling axion field $\phi$ reaches enhanced symmetry points $\phi_i$. At each such value $\phi_i$, a modulus field $\chi_i$ becomes massless \cite{Eva2, Eva1}. 

\subsection{Estimation of the Back-Reaction Effect}

In the Lagrangian eq. (\ref{trappedlagrangian}), the $\chi_i$ fields become massless at the enhanced symmetry points $\phi_i$, which are expected to be evenly spaced in $\phi$. We denote the interval between enhanced symmetry points by
\be
\Delta \, \equiv \, \phi_{i+1}-\phi_i \, .
\ee
Similar to eq. (\ref{background}) and eq. (\ref{backreaction}), the background solution of this model satisfies
\begin{equation}\label{backgroundtrapped}
    \ddot{\phi}_0 + 3H\dot{\phi}_0 + V'(\phi_0) \, = \, 0 \, ,
\end{equation}
while the full solution including the back-reaction effects of the $\chi_i$ fluctuations produced when $\phi$ crosses the enhanced symmetry points is
\begin{equation}\label{backreactiontrapped}
    \ddot{\phi} + 3H\dot{\phi} + V'(\phi) - \sum_i g^2|\phi-\phi_i|\langle\chi_i^2\rangle \, = \, 0.
\end{equation}

Following \cite{KLS97}, the back-reaction term $g^2|\phi-\phi_i|\langle\chi_i^2\rangle$ at the $i$-th crossing can be approximated as $gn_{\chi_i}$, where $n_{\chi_i}$ is the number density of the $\chi_i$ particles. As discussed in \cite{trapping} and \cite{trapped}, particle production happens when the adiabaticity condition is violated in the equation of motion of $\chi_i$, which happens when 
\be
|\phi-\phi_i| \, \lesssim \, \delta\phi \, \equiv \, \sqrt{\frac{|\dot{\phi}(t_i)|}{g}} \, ,
\ee
and only for spatial Fourier modes $\chi_{ik}$ satisfying 
\be
\frac{k^2}{g|\dot{\phi}(t_i)|} \, \lesssim \, 1 \, .
\ee
We will show in eq. (\ref{slowroll}) that $\dot{\phi}$ is constant during inflation under the slow-roll approximation, so that $\delta \phi$ is nearly the same for all of the enhanced symmetry points. The particles produced at the i'th crossing are later diluted by the expansion of space, with a dilution factor $\frac{a(t_i)^3}{a(t)^3}$. As derived in \cite{trapping}, the full computation of the particle production for the $i$-th crossing gives
\begin{equation}
    n_{\chi_{ik}} \, = \, e^{-\pi \frac{k^2}{g|\dot{\phi}(t_i)|}}\frac{a(t_i)^3}{a(t)^3},
\end{equation}
for particles of a specific Fourier mode $k$, and
\begin{equation}\label{nchi}
    n_{\chi_i} \, = \, \int\frac{d^3k}{(2\pi)^3}n_{\chi_{ik}} \, = \, \frac{g^{\frac{3}{2}}}{(2\pi)^3}(\dot{\phi}(t_i))^{\frac{3}{2}}\frac{a(t_i)^3}{a(t)^3}
\end{equation}
for the total number of $\chi$ particles. To obtain a qualitative understanding of this result, note that the back-reaction effect is dominated by the largest value of $k$ for which the mode functions undergo amplification, because the phase space of modes grows as $k^3$. Thus, it is the critical value $k\sim \sqrt{g|\dot{\phi}|}$ which dominates. The time interval of particle production is given by
\be
\Delta t \, \sim \, \frac{\delta \phi}{\dot{\phi}}\sim \frac{1}{\sqrt{g|\dot{\phi}|}} \, .
\ee

\begin{figure}[htbp]
\centering
\includegraphics[scale=0.4]{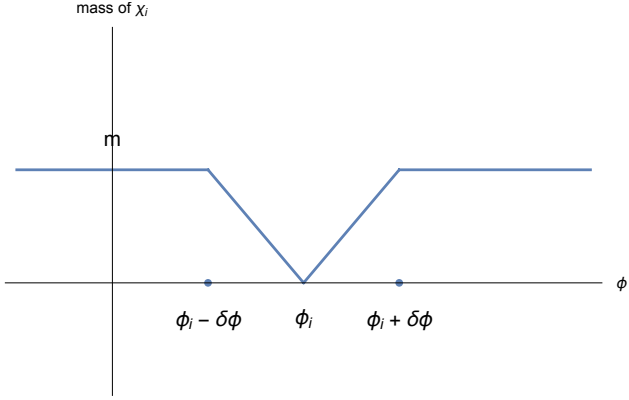}
\caption{The $\chi_i$ mass near the enhanced symmetry point $\phi_i$.}
\label{chimass}
\end{figure}

The Lagrangian (\ref{trappedlagrangian}) cannot be trusted to describe the interaction between $\chi_i$ and $\phi$ for values of $\phi$ which are outside of the interval $|\phi - \phi_i| < \delta \phi$ since the interaction term was based on a quadratic approximation about the enhanced symmetry point. We will hence consider a modified Lagrangian in which
we truncate the $\phi-\chi$ interaction terms to be non-zero only inside the regions of non-adiabatic particle production, and take the mass of $\chi_i$ to be constant outside
\ba\label{trappedlagrangiannew}
    \mathcal{L} \, &=& \, \frac{1}{2}\partial_{\mu}\phi \partial^{\mu}\phi - V(\phi) \\
    &+& \frac{1}{2}\sum_i (\partial_{\mu}\chi_i\partial^{\mu}\chi_i-g^2(\phi-\phi_i)^2\chi_i^2\Theta(\delta\phi-|\phi-\phi_i|)) \nonumber \\
    &-& \frac{1}{2}\sum_i m^2\chi_i^2\Theta(|\phi-\phi_i|-\delta\phi) \, , \nonumber
\ea
where $\Theta(x)$ is the step function, and $m \equiv g\delta\phi$ is the $\chi$ mass outside the non-adiabatic regions. In Figure \ref{chimass}, we show the evolution of the $\chi_i$ mass during the rolling of $\phi$.
Therefore, the equation of motion for $\phi$ becomes
\ba
    \ddot{\phi} &+& 3H\dot{\phi} + V'(\phi) \\
    &-& \sum_i g^2|\phi-\phi_i|\langle\chi_i^2\rangle\Theta(\delta\phi-|\phi-\phi_i|) \, = \, 0 \, ,
    \nonumber
\ea
which is equivalent to
\begin{equation}\label{backreactionnew}
    \ddot{\phi} + 3H\dot{\phi} + V'(\phi) - \sum_i gn_{\chi_i}\Theta(\delta\phi-|\phi-\phi_i|) \, = \, 0 \, .
\end{equation}
The back-reaction terms in eq. (\ref{backreactionnew}) are multiplied by the additional step functions and thus differ from eq. (\ref{backreactiontrapped}). In our model, the regions of production of $\chi_i$ particles lead overlaps of a number ${\tilde{N}}$ of $\phi$ intervals where particle production of different $\chi_i$ fields occurs. The number is given by 
\be
{\tilde{N}} \, \simeq \, \frac{2\delta\phi}{\Delta} \, \simeq \, \frac{2\sqrt{|\dot{\phi}|/g}}{\Delta} \, .
\ee
Combining this result with (\ref{nchi}), the summation of the back-reaction terms can be approximated as
\begin{equation}
    \sum_i gn_{\chi_i}\Theta(\delta\phi-|\phi-\phi_i|) \, \simeq \, \frac{2g^2}{(2\pi)^3}\frac{|\dot{\phi}|^2}{\Delta},
\end{equation}
where we neglected the expansion of space in (\ref{nchi}) since at any time it will be new instability bands which dominate the back-reaction. The effective equation of motion of the $\phi$ field thus becomes
\be\label{chibackreaction}
    \ddot{\phi} + 3H\dot{\phi} + V'(\phi) - \frac{2g^2}{(2\pi)^3}\frac{|\dot{\phi}|^2}{\Delta}
    \, = \, 0 
\ee
with
\be \label{Hubble}
H \, = \, \frac{1}{\sqrt{3}m_{pl}}\sqrt{\frac{1}{2}\dot{\phi}^2+V(\phi)} \, .
\ee
We see that the amplitude of the $\chi$ back-reaction term is determined by a free parameter 
\be
F \, \equiv \, \frac{g^2}{\Delta} \, .
\ee 

We have assumed in eq. (\ref{Hubble}) that the energy density of the $\chi$ particles is sub-dominant compared to $V(\phi)$
\begin{widetext}
\begin{equation}\label{subdominant}
    \rho_\chi \, \simeq \, \sum_i g\delta\phi n_{\chi_i} \, \simeq \,
    \sum_i g\sqrt{\frac{|\dot{\phi}(t_i)|}{g}}\frac{g^{\frac{3}{2}}}{(2\pi)^3}\dot{\phi}(t_i)^{\frac{3}{2}}\frac{a(t_i)^3}{a(t)^3} \,
    \simeq \, \frac{g^2}{3H\Delta(2\pi)^3}|\dot{\phi}(t)|^3 \, \ll \, V(\phi),
\end{equation}
\end{widetext}
so that it is negligible in the Friedmann equation. If we assume slow roll $|\ddot{\phi}|\ll 3H|\dot{\phi}| \ll V'$, the equation (\ref{chibackreaction}) can be solved as
\begin{equation}\label{slowroll}
    |\dot{\phi}| \, \simeq \, \left(\frac{(2\pi)^3 V'\Delta}{2g^2}\right)^{\frac{1}{2}}.
\end{equation}
Substituting eq. (\ref{slowroll}) into eq. (\ref{subdominant}), and using the approximation $V(\phi) \simeq \mu^3 m_{pl}$, we get a constraint on $H$, $F$ and $\mu$
\begin{equation}\label{addrequire}
    H^{-1}F^{-\frac{1}{2}}\mu^{\frac{3}{2}} \, \ll \,  m_{pl}.
\end{equation}

Using eq. (\ref{slowroll}) and assuming $\dot{\phi}^2\ll V(\phi)$ and $\phi(t_{end}) \simeq 0$, the number $N$ of e-foldings of the expansion of space is approximately'
\begin{equation}\label{Napprox}
    N \, \equiv \, \int_0^{t_{end}} Hdt \, \simeq \, \frac{2\sqrt{2}m_{pl}^{\frac{1}{2}}}{3\sqrt{3}(2\pi)^{\frac{3}{2}}}F^{\frac{1}{2}}.
\end{equation}

\subsection{Numerical Solutions and Observational Constraints}

\begin{widetext}
\begin{figure}[htbp]
\centering
\includegraphics[scale=1]{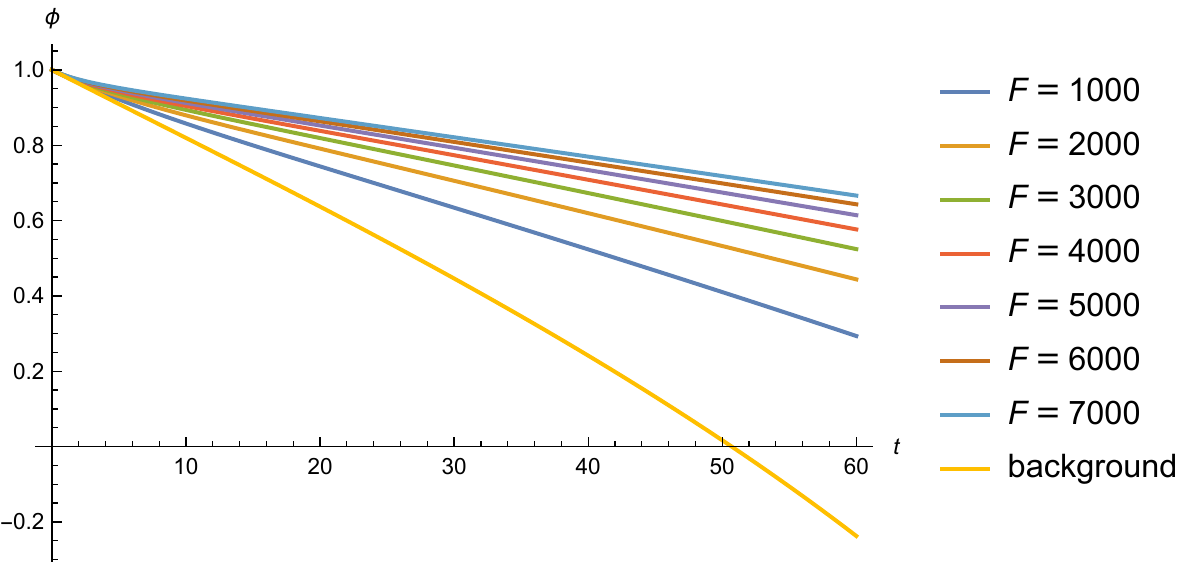}
\caption{The evolution of $\phi$ starting from $\phi=1 m_{pl}$ for different values of $F$, with fixed $\mu = 0.1$ (in Planck units), and compared with the background solution.}
\label{trappedFscan}
\end{figure}
\end{widetext}

In Figure \ref{trappedFscan}, we show the numerical solution of eq. (\ref{chibackreaction}) for various values of the parameter $F$, but fixed $\mu$, and compare the results with the background solution from eq. (\ref{backgroundtrapped}). In contrast to the results for the previous model (where only modulation effects but no $\chi$ particle production was considered) shown in Figure \ref{axionevolution}, in the model considered in this section we see that slow rolling of $\phi$ starting with Planckian initial field value can be realized provided that the parameter $F$ is large enough. 

Furthermore, we checked for which values of our model parameters we can get a sufficient number of e-foldings of accelerated expansion, fixing the initial value of $\phi$ to be
$\phi = m_{pl}$ in order to satisfy the distance criterion, but varying the parameters $\mu$ and $F$, in order to check whether there exists a range satisfying both of the criteria. Note that for field values $|\phi| < m_{pl}$ the de Sitter constraint is automatically satisfied. 

In order to obtain an inflationary model consistent with current observational constraints, we will fix the inflationary slow-roll parameters to be \cite{Planck} (at the beginning of the period of inflation)
\ba
    \epsilon \, &\equiv& \, -\frac{\dot{H}}{H^2} \, \simeq \, 0.01 \,\,\, {\rm{and}} \nonumber \\  A_s \, &\equiv& \, \frac{H^2}{8\pi^2\epsilon m_{pl}^2} \, \simeq \, 2\times 10^{-9} \, ,
\ea
the second condition coming from the amplitude of the spectrum of cosmological perturbations, and the first from its slope. From the $A_s$ (scalar amplitude) constraint and using eq. (\ref{chibackreaction}), we get
\begin{equation}\label{Hconstrain}
    H \, \lesssim \, \frac{1}{\sqrt{3}m_{pl}}\sqrt{\mu^3 m_{pl}} \, \simeq \, 4\times 10^{-5} m_{pl},
\end{equation}
which sets a range for $\mu$:
\begin{equation}\label{muconstrain}
    \mu \, \lesssim \, 0.0017 m_{pl}.
\end{equation}

We also require at least 60 e-foldings of spatial expansion during the whole period of inflation
\begin{equation}
    N \, \geq \,  60 \, ,
\end{equation}
where the end time of inflation is determined by $\epsilon(t_{end}) = 1$. It can also be numerically checked that $\phi(t_{end}) \simeq 0$, $\dot{\phi}^2(t_{end}) \lesssim \mu^3 \phi(t_{end})$, verifying that the potential energy of $\phi$ is larger than the kinetic energy during the period of inflation. Using eq. (\ref{Napprox}), we get the constraint
\begin{equation}\label{Fapprox}
    F \, \simeq \, \frac{27(2\pi)^3N^2}{8m_{pl}}\simeq 3\times 10^6 m_{pl}^{-1}.
\end{equation}
Substituting eq. (\ref{Hconstrain}), eq. (\ref{Fapprox}) and eq. (\ref{muconstrain}) into (\ref{addrequire}), we find that the consistency constraint $\rho_{\chi}\ll\rho_{\phi}$ is satisfied for the chosen parameters.

\begin{widetext}
\begin{figure}[htbp]
\centering
\includegraphics[scale=0.4]{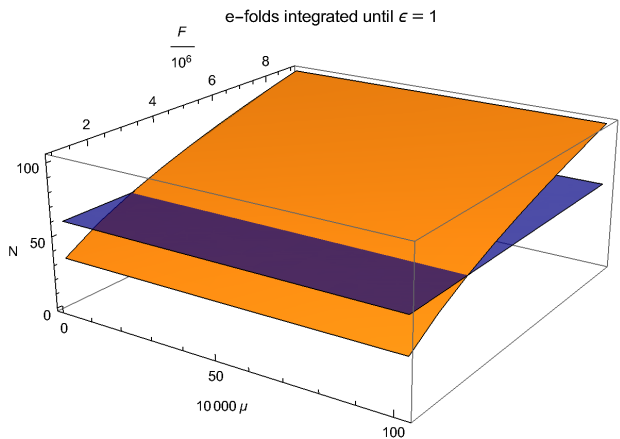}
\caption{The e-folds in a 2D parameter scan of $\mu$ and $F$ in Plank units (orange), compared with the 60 e-folds' surface (blue).}
\label{2Dscan}
\end{figure}
\end{widetext}

In Figure \ref{2Dscan}, we show the results of a scan of the parameter space 
\ba
0 \, &<& \, \mu \lesssim \, 0.01 m_{pl} \\
10^6 m_{pl}^{-1} \, &\leq& \, F \, \leq \, 9\times 10^6 m_{pl}^{-1} \, , \nonumber
\ea
and calculate the number of e-foldings of inflation (which is shown on the vertical axis in orange colour) for each grid point. Comparing the results with the required 60 e-foldings (which is shown as the horizontal surface in blue), we find that inflation consistent with both observations and the swampland constraints can be obtained if the parameter values are close to the intersection line of the two surfaces. The numerical results are compatible with what can be obtained from eq. (\ref{Fapprox}). Note that - if we add the oscillating term in the potential considered in Section II - the value of $\Delta$ is bounded from below by the analysis of Section II. The rough bound is $\Delta > 10^{-3} m_{pl}$. Hence, the requirement of a large value of $F$ from (\ref{Fapprox}) implies that the coupling of $\phi$ to the moduli fields must be strong, i.e. $g^2 > 10^3$. 

To conclude, the axion monodromy model of eq. (\ref{trappedlagrangiannew}) which includes the effects of moduli production at enhanced symmetry points can be consistent with both the slow-roll condition and the de Sitter Conjecture given suitable parameters. This result agrees with the conclusions of \cite{trapping} (see also \cite{Vahid}).

\begin{thebibliography}{99}

\bibitem{inflation}
A.~H.~Guth,
 ``The Inflationary Universe: A Possible Solution to the Horizon and Flatness Problems,''
 Phys.\ Rev.\ D {\bf 23}, 347 (1981)
 [Adv.\ Ser.\ Astrophys.\ Cosmol.\  {\bf 3}, 139 (1987)].
 doi:10.1103/PhysRevD.23.347;\\
 R.~Brout, F.~Englert and E.~Gunzig,
 ``The Creation Of The Universe As A Quantum Phenomenon,''
 Annals Phys.\  {\bf 115}, 78 (1978);\\
 A.~A.~Starobinsky,
 ``A New Type Of Isotropic Cosmological Models Without Singularity,''
 Phys.\ Lett.\ B {\bf 91}, 99 (1980);\\
 K.~Sato,
 ``First Order Phase Transition Of A Vacuum And Expansion Of The Universe,''
 Mon.\ Not.\ Roy.\ Astron.\ Soc.\  {\bf 195}, 467 (1981).
 
\bibitem{Brennan}
T.~D.~Brennan, F.~Carta and C.~Vafa,
 ``The String Landscape, the Swampland, and the Missing Corner,''
 PoS TASI {\bf 2017}, 015 (2017)
 doi:10.22323/1.305.0015
 [arXiv:1711.00864 [hep-th]].
 
\bibitem{Palti}
 E.~Palti,
 ``The Swampland: Introduction and Review,''
Fortsch.\ Phys.\  {\bf 67}, no. 6, 1900037 (2019)
  doi:10.1002/prop.201900037
  [arXiv:1903.06239 [hep-th]].
 
\bibitem{Vafa1}
H.~Ooguri and C.~Vafa, 
``On the Geometry of the String Landscape and the Swampland,'' 
  Nucl.\ Phys.\ B {\bf 766}, 21 (2007)
  [hep-th/0605264].
  
\bibitem{Vafa2}
G.~Obied, H.~Ooguri, L.~Spodyneiko and C.~Vafa,
  ``De Sitter Space and the Swampland,''
  arXiv:1806.08362 [hep-th].
 
\bibitem{Krishnan}
S.~K.~Garg and C.~Krishnan,
  ``Bounds on Slow Roll and the de Sitter Swampland,''
  JHEP {\bf 1911}, 075 (2019)
  doi:10.1007/JHEP11(2019)075
  [arXiv:1807.05193 [hep-th]].
  
\bibitem{noinflation}
P.~Agrawal, G.~Obied, P.~J.~Steinhardt and C.~Vafa,
  ``On the Cosmological Implications of the String Swampland,''
  Phys.\ Lett.\ B {\bf 784}, 271 (2018)
  doi:10.1016/j.physletb.2018.07.040
  [arXiv:1806.09718 [hep-th]].
  
\bibitem{Lavinia}
L.~Heisenberg, M.~Bartelmann, R.~Brandenberger and A.~Refregier,
  ``Dark Energy in the Swampland,''
  Phys.\ Rev.\ D {\bf 98}, no. 12, 123502 (2018)
  doi:10.1103/PhysRevD.98.123502
  [arXiv:1808.02877 [astro-ph.CO]].
  
\bibitem{heisenberg2019horndeski}
   L.~Heisenberg, M.~Bartelmann, R.~Brandenberger, and A.~Refregier,
   ``Horndeski gravity in the swampland,''
   Phys.\ Rev.\ D {\bf 99}, 124020 (2019) doi:10.1103/PhysRevD.99.124020
   [arXiv:1902.03939[hep-th]].
   
\bibitem{Dvali}
G.~Dvali, C.~Gomez and S.~Zell,
  ``Quantum Break-Time of de Sitter,''
  JCAP {\bf 1706}, 028 (2017)
  doi:10.1088/1475-7516/2017/06/028
  [arXiv:1701.08776 [hep-th]].
  
\bibitem{Keshav}
S.~Brahma, K.~Dasgupta and R.~Tatar,
  ``Four-dimensional de Sitter space is a Glauber-Sudarshan state in string theory,''
  arXiv:2007.00786 [hep-th];\\
  S.~Brahma, K.~Dasgupta and R.~Tatar,
  ``de Sitter Space as a Glauber-Sudarshan State,''
  arXiv:2007.11611 [hep-th].
  
\bibitem{warm}
A.~Berera,
  ``Warm inflation,''
  Phys.\ Rev.\ Lett.\  {\bf 75}, 3218 (1995)
  doi:10.1103/PhysRevLett.75.3218
  [astro-ph/9509049].
  
\bibitem{trapped}
 D.~Green, B.~Horn, L.~Senatore and E.~Silverstein,
  ``Trapped Inflation,''
  Phys.\ Rev.\ D {\bf 80}, 063533 (2009)
  doi:10.1103/PhysRevD.80.063533
  [arXiv:0902.1006 [hep-th]].
  
\bibitem{chromo}
P.~Adshead and M.~Wyman,
  ``Chromo-Natural Inflation: Natural inflation on a steep potential with classical non-Abelian gauge fields,''
  Phys.\ Rev.\ Lett.\  {\bf 108}, 261302 (2012)
  doi:10.1103/PhysRevLett.108.261302
  [arXiv:1202.2366 [hep-th]].
  
\bibitem{Bedroya}
  A.~Bedroya and C.~Vafa,
  ``Trans-Planckian Censorship and the Swampland,''
JHEP {\bf 2009}, 123 (2020)
  doi:10.1007/JHEP09(2020)123
  [arXiv:1909.11063 [hep-th]].
  
\bibitem{Bedroya2}
A.~Bedroya, R.~Brandenberger, M.~Loverde and C.~Vafa,
  ``Trans-Planckian Censorship and Inflationary Cosmology,''
  Phys.\ Rev.\ D {\bf 101}, no. 10, 103502 (2020)
  doi:10.1103/PhysRevD.101.103502
  [arXiv:1909.11106 [hep-th]].
  
\bibitem{Vahid}
A.~Berera, R.~Brandenberger, V.~Kamali and R.~Ramos,
  ``Thermal, Trapped and Chromo-Natural Inflation in light of the Swampland Criteria and the Trans-Planckian Censorship Conjecture,''
  arXiv:2006.01902 [hep-th].

\bibitem{heisenberg2020model}
L.~Heisenberg, M.~Bartelmann, R.~Brandenberger and A.~Refregier,
``Model Independent Analysis of Supernova Data, Dark Energy, Trans-Planckian Censorship and the Swampland,''
arXiv:2003.13283 [hep-th].


\bibitem{Liam}
D.~Baumann and L.~McAllister,
  ``Inflation and String Theory,''
  doi:10.1017/CBO9781316105733
  arXiv:1404.2601 [hep-th].
  
\bibitem{Eva1}
E.~Silverstein and A.~Westphal,
  ``Monodromy in the CMB: Gravity Waves and String Inflation,''
  Phys.\ Rev.\ D {\bf 78}, 106003 (2008)
  doi:10.1103/PhysRevD.78.106003
  [arXiv:0803.3085 [hep-th]].
  
\bibitem{Eva2}
L.~McAllister, E.~Silverstein and A.~Westphal,
  ``Gravity Waves and Linear Inflation from Axion Monodromy,''
  Phys.\ Rev.\ D {\bf 82}, 046003 (2010)
  doi:10.1103/PhysRevD.82.046003
  [arXiv:0808.0706 [hep-th]].
  
\bibitem{trapping}
L.~Kofman, A.~D.~Linde, X.~Liu, A.~Maloney, L.~McAllister and E.~Silverstein,
  ``Beauty is attractive: Moduli trapping at enhanced symmetry points,''
  JHEP {\bf 0405}, 030 (2004)
  doi:10.1088/1126-6708/2004/05/030
  [hep-th/0403001].
  
\bibitem{Scott}
S. Watson,
  ``Moduli stabilization with the string Higgs effect,''
  Phys.\ Rev.\ D {\bf 70}, 066005 (2004)
  doi:10.1103/PhysRevD.70.066005
  [hep-th/0404177].
  
\bibitem{preheating}
A.~D.~Dolgov and D.~P.~Kirilova,
  ``On Particle Creation By A Time Dependent Scalar Field,''
  Sov.\ J.\ Nucl.\ Phys.\  {\bf 51}, 172 (1990)
  [Yad.\ Fiz.\  {\bf 51}, 273 (1990)];\\
  J.~H.~Traschen and R.~H.~Brandenberger,
  ``Particle Production During Out-of-equilibrium Phase Transitions,''
  Phys.\ Rev.\ D {\bf 42}, 2491 (1990)
  doi:10.1103/PhysRevD.42.2491;\\
  L.~Kofman, A.~D.~Linde and A.~A.~Starobinsky,
  ``Reheating after inflation,''
  Phys.\ Rev.\ Lett.\  {\bf 73}, 3195 (1994)
  doi:10.1103/PhysRevLett.73.3195
  [hep-th/9405187];\\
  Y.~Shtanov, J.~H.~Traschen and R.~H.~Brandenberger,
  ``Universe reheating after inflation,''
  Phys.\ Rev.\ D {\bf 51}, 5438 (1995)
  doi:10.1103/PhysRevD.51.5438
  [hep-ph/9407247];\\
  R.~Allahverdi, R.~Brandenberger, F.~Y.~Cyr-Racine and A.~Mazumdar,
  ``Reheating in Inflationary Cosmology: Theory and Applications,''
  Ann.\ Rev.\ Nucl.\ Part.\ Sci.\  {\bf 60}, 27 (2010)
  doi:10.1146/annurev.nucl.012809.104511
  [arXiv:1001.2600 [hep-th]];\\
  M.~A.~Amin, M.~P.~Hertzberg, D.~I.~Kaiser and J.~Karouby,
  ``Nonperturbative Dynamics Of Reheating After Inflation: A Review,''
  Int.\ J.\ Mod.\ Phys.\ D {\bf 24}, 1530003 (2014)
  doi:10.1142/S0218271815300037
  [arXiv:1410.3808 [hep-ph]].
 
\bibitem{KLS97}
  L.~Kofman, A.~D.~Linde and A.~A.~Starobinsky,
  ``Towards the theory of reheating after inflation,''
  Phys.\ Rev.\ D {\bf 56}, 3258 (1997)
  doi:10.1103/PhysRevD.56.3258
  [hep-ph/9704452].
 
\bibitem{Planck}
 N.~Aghanim {\it et al.} [Planck Collaboration],
  ``Planck 2018 results. VI. Cosmological parameters,''
  Astron.\ Astrophys.\  {\bf 641}, A6 (2020)
  doi:10.1051/0004-6361/201833910
  [arXiv:1807.06209 [astro-ph.CO]].
 
\bibitem{Vahid2}
   V.~Kamali and R.~Brandenberger,
  ``Relaxing the TCC Bound on Inflationary Cosmology?,''
  Eur.\ Phys.\ J.\ C {\bf 80}, no. 4, 339 (2020)
  doi:10.1140/epjc/s10052-020-7908-8
  [arXiv:2001.00040 [hep-th]].
  
\bibitem{Dvali2}
G.~Dvali, A.~Kehagias and A.~Riotto,
  ``Inflation and Decoupling,''
  arXiv:2005.05146 [hep-th].
  

\end{thebibliography}

\section{Conclusions}

We have studied the evolution of the axion field $\phi$ in an axion monodromy model without and with the consideration of moduli production at enhanced symmetry points. Without considering the production of such moduli fields, the production of $\phi$ particles due to the nonlinearities in the $\phi$ potential which is induced by the monodromy effects is unable to render the model consistent with the swampland criteria. Either the particle production effects are too weak to significantly affect the background dynamics, or else they lead to trapping in a local minimum and thus to a {\it graceful exit} problem. On the other hand, moduli production at enhanced symmetry points is able to slow down (for a certain range of the model parameters) the background field dynamics sufficiently such that a long enough period of inflation starting with field values consistent with the distance conjecture is possible. The model is then also consistent with the de Sitter conjecture. The model which emerges is, however, not standard slow roll inflation, but rather a {\it trapped inflation} scenario.

There is a connection between our study and the arguments  \cite{noinflation} based on the swampland criteria that minimal slow-roll inflation might not be realizable in string theory (naturally, it is also possible that string theory may provide a counterexample to the swampland conjectures). In the example we have studied here, string theory considerations transform a model which at first sight looks as if it could yield a realization of the standard slow roll scenario into a model which is different and which is consistent with the swampland criteria. The resulting model, however, is still in tension with the TCC, which states \cite{Bedroya} that no length scale which was initially smaller than the Planck length is allowed to become super-Hubble. This sets an upper bound on the the duration of inflation. Demanding that inflation can explain the origin of structure on all currently observed scales sets a lower bound on the duration of inflation. These bounds are only consistent \cite{Bedroya2} if the energy scale of inflation is lower than about $\eta \sim 10^{9} {\rm{GeV}}$ (for power law inflation the bound is somewhat relaxed \cite{Vahid2}). Models with such a low energy scale require severe fine tuning in order to be consistent with observations \cite{Bedroya2}. Once again, however, it is possible that the TCC criteria can be violated in the context of a more consistent theory of the early universe (see e.g. \cite{Dvali2}).

\section*{Acknowledgements}

We are grateful to Liam McAllister for feedback on the manuscript.
RB thanks the  Pauli Center and the Institutes of Theoretical Physics and of Particle- and Astrophysics of the ETH for hospitality. The research at McGill is supported, in part, by funds from NSERC and from the Canada Research Chair program. LH is supported by funding from the European Research Council (ERC) under the European Unions Horizon 2020 research and innovation programme grant agreement
No 801781 and by the Swiss National Science Foundation grant 179740.

\end{document}